\documentclass[10pt,aps,showpacs,floatfix,twocolumn,tightenlines]{revtex4}
\usepackage{epsfig}
\usepackage{psfig}
\usepackage{graphicx}
\usepackage{graphics}
\usepackage{xspace}
\usepackage{amssymb}
\usepackage{amsmath}
\usepackage{latexsym}
\usepackage{natbib}
\usepackage{mathrsfs}
\newcommand{\inieq}{\begin{eqnarray}}            
\newcommand{\fineq}{\end{eqnarray}}            
\newcommand{\diff}{{\rm\,d}}                    
\newcommand{\bint}{\mskip .5mu \int \mskip-18mu} 

\def\p{\mbox{\boldmath $p$}}

\def\q{\mbox{\boldmath $q$}}

\def\k{\mbox{\boldmath $k$}}

\def\mcg{\mbox{$\mathcal{G}$}}
\def\mcv{\mbox{$\mathcal{V}$}}
\begin{document}
\title{Relativistic Green's function approach to charged-current
neutrino-nucleus quasielastic scattering} 

\author{Andrea Meucci} 
\author{Carlotta Giusti}
\author{Franco Davide Pacati }
\affiliation{Dipartimento di Fisica Nucleare e Teorica, 
Universit\`{a} degli Studi di Pavia and \\
Istituto Nazionale di Fisica Nucleare, 
Sezione di Pavia, I-27100 Pavia, Italy}

\date{\today}

\begin{abstract}
A relativistic Green's function approach to inclusive quasielastic 
charged-current neutrino-nucleus scattering is developed. 
The components of the hadron tensor are written in terms of the single-particle
Green's function, which is expanded on the eigenfunctions 
of the nuclear optical potential, so that final state 
interactions are accounted for by means of a complex optical potential but 
without a loss of flux. 
Results for the $(\nu_{\mu},\mu^-)$ reaction on $^{16}$O and $^{12}$C 
target nuclei are presented and discussed. A reasonable agreement of 
the flux-averaged cross section on $^{12}$C with experimental data is achieved. 
\end{abstract}
\pacs{25.30.Pt: Neutrino scattering, 13.15.+g: Neutrino interactions, 
24.10.Jv: Relativistic models, 24.10.Cn: Many-body theory}

\maketitle

\section{Introduction}

The reactions with an incident neutrino interacting with a complex nucleus 
have gained 
in recent years a wide interest, owing both to astrophysical reasons and to the
aim of investigating the neutrino properties with a high accuracy.
Besides the measurements with large underground detectors,
some experiments have also been performed~\cite{albert,athan,auer} using a
pion beam which weakly decays producing leptons. 
In this case the most part of the neutrinos which are obtained is related to 
muons, with a smaller component of electron neutrinos.

Both weak neutral- and charged-current scattering have stimulated detailed 
investigations. 
In particular, we are here interested in charged-current reactions at an 
energy below 1 GeV as they have shown to be dependent on nuclear structure 
effects. 
Different approaches have been applied to investigate such processes, 
including the so-called \lq\lq elementary particle model\rq\rq~\cite{min},  
random phase approximation (RPA) in the framework of a relativistic Fermi gas 
model~\cite{kim} or a Fermi gas model with local density 
approximation~\cite{singh}, shell model~\cite{hayes,volpe} and relativistic 
shell model~\cite{alberico}, RPA among quasiparticles~\cite{volpe} and 
continuum RPA~\cite{kolbe,jacho}. 
The reaction goes through a quasielastic (QE) mechanism, where the neutrino 
interacts with one single neutron and a proton together with a negative muon 
are emitted.
The effect of final state interactions (FSI) has been stressed to significantly 
contribute to the cross section~\cite{bleve} and has been calculated with a 
relativistic shell model including a phenomenological optical potential, which 
describes the interaction of the outgoing nucleon with the residual nucleus, 
with and without imaginary part~\cite{maieron}.

The optical potential is fitted to reproduce the elastic proton-nucleus 
scattering through its real component, while its imaginary part takes into
account the scattering towards the inelastic channels. 
This means that the reaction channels are globally described by
a loss of flux produced by the imaginary part of the complex potential.
This model has been applied with great success to exclusive QE electron 
scattering~\cite{book}, where it is able to explain the experimental cross
sections of one-nucleon knockout reactions in a range of nuclei from carbon to 
lead. In an inclusive process, however, where some of the reaction products are 
not detected and the inelastic channels are also included in the experimental 
cross section, the flux must be conserved.
This fact is sometimes described by dropping the imaginary part of the optical 
potential. This procedure conserves the flux but it is not consistent with 
the exclusive reaction, which can only be reproduced with a careful treatment 
of the optical potential, including both real and imaginary parts~\cite{book}.

In this paper we apply a Green's function approach where the conservation of 
flux is preserved and FSI are treated in the inclusive reaction consistently 
with the exclusive one. This method was discussed in a 
nonrelativistic~\cite{capuzzi} and in a relativistic~\cite{ee} framework for 
the case of inclusive electron scattering and it is here adapted, in a 
relativistic framework, to charged-current neutrino scattering.
In this approach, the components of the nuclear response are written in terms 
of the single-particle optical-model Green's function. This result was 
originally derived by arguments based on the multiple scattering 
theory~\cite{hori} and successively by means of the Feshbach projection 
operator formalism~\cite{capuzzi,chinn,bouch,capma}. The spectral 
representation of the single-particle Green's function, based on a biorthogonal 
expansion in terms of the eigenfunctions of the non-Hermitian optical 
potential, allows one to perform explicit calculations and to treat FSI 
consistently in the inclusive and in the exclusive reactions. Important 
and peculiar effects are given in the inclusive ($e,e'$) reaction by the 
imaginary part of the optical potential, which is responsible for the  
redistribution of the strength among different channels. 
 
In Sec. II the general formalism of the charged-current neutrino-nucleus
scattering is given. In Sec. III, the Green's function 
approach is briefly reviewed. In Sec. IV, the results obtained on $^{16}$O and 
$^{12}$C target nuclei are presented and discussed. Some conclusions are drawn 
in Sec. V.

\section{The inclusive cross section}
\label{sec.cross}

In a charged-current process neutrinos and antineutrinos interact with nuclei
via the exchange of weak-vector bosons and charged leptons are produced in the
final state.
The cross section of an inclusive reaction where an incident neutrino or
antineutrino, with four-momentum $k^\mu_i = (\varepsilon_i,\k_i)$, is absorbed
by a nucleus and only the outgoing lepton, with four-momentum 
$k^\mu = (\varepsilon,\k)$ and mass $m$, is detected, is given by the contraction 
between the lepton tensor and the hadron tensor, i.e.,
\begin{eqnarray}
\diff \sigma = \frac {G^2 \cos^2\vartheta_{\textrm c}} {2} \ 2\pi \
 L^{\mu\nu}\ W_{\mu\nu}\ \frac {\diff^3k} {(2\pi)^3} \ ,
\label{eq.cs1}
\end{eqnarray}
where $G \simeq 1.16639 \times 10^{-11}$ MeV$^{-2}$ is the Fermi constant 
and $\vartheta_{\textrm c}$ is the Cabibbo angle 
($\cos \vartheta_{\textrm c}\simeq 0.9749$). 

The lepton tensor is 
\begin{eqnarray}
L^{\mu\nu} = \frac {1} {2\varepsilon_i} \frac {m} {\varepsilon} 
\sum_{\mathrm {spin}}
\bar{u}_f \gamma^\mu (1\mp \gamma^5) u_i \ \bar{u}_i (1\pm \gamma^5) \gamma^\nu
u_f,
\label{eq.lt}
\end{eqnarray}
where the upper (lower) sign corresponds to neutrino
(antineutrino) scattering. 
After projecting into the initial neutrino (antineutrino) and the final lepton 
state, one has
\begin{eqnarray}
L^{\mu\nu} = \frac {1} {2\varepsilon_i \varepsilon} 
\mathrm{Tr} \left[ \gamma \cdot k \ \gamma^\mu \ (1\mp \gamma^5) \ 
\gamma \cdot k_i \ \gamma^\nu \right],
\label{eq.lt1}
\end{eqnarray}
which can be written, by separating the symmetrical and antisymmetrical 
components, as 
\begin{eqnarray}
L^{\mu\nu} = \frac {2} {\varepsilon_i \varepsilon} 
\left[ l_S^{\mu\nu} \mp l_A^{\mu\nu} \right],
\label{eq.lt2}
\end{eqnarray}
with
\begin{eqnarray}
l_S^{\mu\nu} &=& k_i^\mu \ k^\nu + k_i^\nu \ k^\mu - g^{\mu\nu} \ k_i \cdot k
\nonumber \\
l_A^{\mu\nu} &=& i \ \epsilon ^{\mu\nu\alpha\beta} k_{i\alpha} k_\beta ,
\label{eq.lt3}
\end{eqnarray}
where $\epsilon ^{\mu\nu\alpha\beta}$ is the antisymmetric tensor with 
$\epsilon_{0123} = - \epsilon^{0123} = 1$.

Assuming the reference frame where the $z$-axis is parallel to the 
momentum transfer $\q = \k_i - \k$ and the $y$-axis is parallel to 
$\k_i \times \k$, the symmetrical components 
$l_S^{0y}, l_S^{xy}, l_S^{zy}$, and the antisymmetrical ones
$l_A^{0x}, l_A^{xz}, l_A^{0z}$, as well as those obtained from 
them by exchanging their indices, vanish.

The hadron tensor is given by bilinear products of the transition matrix
elements of the nuclear weak charged-current operator $J^{\mu}$ between
the initial state $\mid\Psi_0\rangle$ of the nucleus, of energy $E_0$, and the 
final states $\mid \Psi_{\textrm {f}} \rangle$, of energy $E_{\textrm {f}}$, 
both eigenstates of the $(A+1)$-body Hamiltonian $H$, as 
\begin{eqnarray}
& & W^{\mu\nu}(\omega,q) = 
 \bint\sum_{\textrm {f}}  \langle 
\Psi_{\textrm {f}}\mid J^{\mu}(\q) \mid \Psi_0\rangle \nonumber \\
&\times&  \langle 
\Psi_0\mid J^{\nu\dagger}(\q) \mid \Psi_{\textrm {f}}\rangle 
\ \delta (E_0 +\omega - E_{\textrm {f}}).
\label{eq.ha1}
\end{eqnarray}
and involves an average over initial states and a sum over the undetected final 
states. The sum runs over the scattering states corresponding to all of the 
allowed asymptotic configurations and includes possible discrete states.  

The transition matrix elements are calculated in the first
order perturbation theory and in the impulse approximation, i.e., the
incident neutrino interacts with only one nucleon while the other ones behave 
as spectators. The current operator is assumed to be adequately described as 
the sum of single-nucleon currents, corresponding to the weak charged current  
\begin{eqnarray}
  j^{\mu} &=& \big[ F_1^{\textrm V}(Q^2) \gamma ^{\mu} + 
             i\frac {\kappa}{2M} F_2^{\textrm V}(Q^2)\sigma^{\mu\nu}q_{\nu} 
	     \nonumber \\ 
	     &-&G_{\textrm A}(Q^2)\gamma ^{\mu}\gamma ^{5} +
	     F_{\textrm P}(Q^2)q^{\mu}\gamma ^{5} \big] \tau^{\pm},
	     \label{eq.cc}
\end{eqnarray}
where $\tau^{\pm}$ are the isospin operators, $\kappa$ is the anomalous part of 
the magnetic moment, $q^{\mu} = (\omega , \q)$, with $Q^2 = |\q|^2 - \omega^2$, 
is the four-momentum transfer, and
$\sigma^{\mu\nu}=\left(i/2\right)\left[\gamma^{\mu},\gamma^{\nu}\right]$.
$F_1^{\textrm V}$ and $F_2^{\textrm V}$ are the isovector Dirac and Pauli 
nucleon form factors, which are taken from Ref.~\cite{mmd}. 
$G_{\textrm A}$ and $F_{\textrm P}$ are the axial and 
induced pseudoscalar form factors, which are usually parametrized as
\begin{eqnarray}
G_{\textrm A} &=& \frac{g_{\textrm A} }
        {\left(1+Q^2/M^2_{\textrm A}\right)^2} \ , \\ 
F_{\textrm P}&=& \frac{2MG_{\textrm A}}{m^2_{\pi} + Q^2} \ , \label{eq.formf}
\end{eqnarray}
where $g_{\textrm A} = 1.267$, $m_{\pi}$ is the pion mass, 
and $M_{\textrm A} \simeq 1.032$ GeV is the axial mass. 

The most general covariant form of the hadron tensor is obtained in terms 
of the two independent four-vectors of the problem, i.e. the four-momentum
transfer $q^\mu$ and the four-momentum $P^\mu$ of the target.
The symmetrical and antisymmetrical components of the tensor can therefore 
be written as
\begin{eqnarray}
W^{\mu\nu}_S &=& W_1\ g^{\mu\nu} + W_2 \ q^\mu q^\nu + W_3 \ P^\mu P^\nu
\nonumber \\
&+& W_4 \ (P^\mu q^\nu + q^\mu P^\nu) \nonumber \\
W^{\mu\nu}_A &=& W_5 \ (P^\mu q^\nu - q^\mu P^\nu) + W_6
\ \epsilon^{\mu\nu\alpha\beta} q_\alpha P_\beta ,
\label{eq.inv}
\end{eqnarray}
where the invariant form factors $W_i$ are functions of the two scalars 
which can be formed with $q^\mu$ and $P^\mu$, i.e., $Q^2$ and $P \cdot q$.
From Eq. (\ref{eq.inv}), it is clear that in our reference frame 
$W^{0x}, W^{0y}, W^{xz}, W^{yz}$, and $W^{xy}_S$ 
vanish together with the tensor components obtained from them by 
exchanging their indices.

The inclusive cross section for the QE $\nu$($\bar\nu$)-nucleus 
scattering, obtained from the contraction between the lepton and hadron 
tensors, can therefore be written as \cite{Walecka}
\begin{eqnarray}
\frac{\diff \sigma} {\diff \varepsilon \diff \Omega} &=& 
 k\varepsilon \ \frac{G^2} {4 \pi^2}
  \cos^2\vartheta_{\textrm c} 
  \Big[ v_0 R_{00} + v_{zz} R_{zz}  \nonumber \\  &-&
  \ v_{0z} R_{0z} + v_T R_T \pm v_{xy} R_{xy} \Big]   \ .
\label{eq.cs}
\end{eqnarray} 
The coefficients $v$, obtained from the lepton tensor components, are
\begin{eqnarray}
v_0 &=& 1 + \tilde k \cos\vartheta \ , \nonumber \\
v_{zz} &=& 1 + \tilde k \cos\vartheta - 2 \frac{\varepsilon_i |\k|
\tilde k} {|\q|^2} \sin^2\vartheta \ , \nonumber \\
v_{0z}&=& \frac{\omega}{|\q|} \left(1 + \tilde k \cos\vartheta\right) + \frac
                    {m^2}{|\q|\varepsilon} \ , \nonumber \\
v_T&=& 1 - \tilde k \cos\vartheta + \frac{\varepsilon_i |\k|\tilde k} {|\q|^2}
                   \sin^2\vartheta \ , \nonumber \\
v_{xy}&=& \frac {\varepsilon_i + \varepsilon}{|\q|} \big( 1 - \tilde k 
\cos\vartheta \big) -
\frac {m^2}{|\q|\varepsilon} \ , 
\label{eq.lepton}
\end{eqnarray}
where $\tilde k = |\k|/\varepsilon$, $\vartheta$ is the lepton
scattering angle, and $m$ the mass of the emitted lepton, e.g., the muon mass
$\simeq 105.9$ MeV. It was taken
\begin{eqnarray}
v_T = \frac {1} {2\varepsilon_i \varepsilon} (l^{xx}_S + l^{yy}_S ),
\end{eqnarray}
taking advantage of the fact that $W^{xx} = W^{yy}$, as can be deduced from 
Eq. (\ref{eq.inv}).

The response functions are given in terms of the components of the 
hadron tensor as
\begin{eqnarray}
R_{00} &=& W^{00}\ , \nonumber \\
R_{zz} &=& W^{zz}\ , \nonumber \\
R_{0z} &=& W^{0z} + W^{z0}\ , \nonumber \\
R_T  &=& W^{xx} + W^{yy}\ , \nonumber \\
R_{xy} &=& i\left( W^{yx} - W^{xy}\right)\ .
\label{eq.rf}
\end{eqnarray}

\section{The relativistic Green's function approach}
\label{sec.green}

All nuclear structure information is contained in the response
functions and therefore in the hadron tensor. We apply here to the inclusive 
QE $\nu$($\bar\nu$)-nucleus scattering the same relativistic 
approach~\cite{ee} which was rather successful in reproducing the cross sections
of the inclusive QE electron scattering. Here we recall only the main features
of the model. More details can be found in Refs.~\cite{ee,book,capuzzi} 

For the inclusive process the components of the hadron tensor in 
Eq. (\ref{eq.ha1}) can equivalently be expressed as 
\inieq
 W^{\mu\nu}(\omega,q)  
=-\frac{1}{\pi} \textrm{Im} \langle \Psi_0
\mid J^{\nu\dagger}(\q) G(E_{\textrm {f}}) J^{\mu}(\q) \mid \Psi_0 \rangle \ , 
\label{eq.hadrontensor}
\fineq
in terms of the Green's function $G(E_{\textrm {f}})$ related to the nuclear 
Hamiltonian $H$, i.e.,
\inieq
G(E_{\textrm {f}}) = \frac{1}{E_{\textrm {f}} - H + i\eta} \ . \label{eq.green}
\fineq
Here and in all the equations involving $G$ the limit for $\eta \rightarrow
+0$ is understood. It must be performed after calculating the matrix elements 
between normalizable states. 

The current operator is assumed to be adequately described as the sum 
of the single-nucleon currents of Eq. (\ref{eq.cc}).

It was shown in Refs. \cite{book,capuzzi,ee,hori} for the inclusive ($e,e'$)
scattering that the components of the nuclear
response in Eq. (\ref{eq.hadrontensor}) can be written in terms of the
single-particle  Green's function $\mcg(E)$, whose self-energy is the 
Feshbach's optical potential. An explicit calculation of the Green's function
can be avoided by its spectral representation, which is based on a biorthogonal 
expansion in terms of the eigenfunctions of the non-Hermitian optical potential
$\mcv$, and of its Hermitian conjugate $\mathcal{V}^{\dagger}$, i.e.,
\inieq
\left[ {\mathcal{E}} - T - {\mathcal{V}}^{\dagger} (E) \right] \mid
{\chi}_{\mathcal{E}}^{(-)}(E)\rangle &=& 0, \nonumber \\
\left[ \mathcal{E} - T - {\mathcal{V}}(E) \right] \mid \tilde
{\chi}_{\mathcal{E}}^{(-)}(E)\rangle &=& 0.
 \label{eq.op}
\fineq
Note that $E$ and ${\mathcal{E}}$ are not necessarily the same.
The spectral representation is
\inieq
\mcg(E) &=& \int_M^{\infty} \diff \mathcal{E}\mid\tilde
{\chi}_{\mathcal{E}}^{(-)}(E)\rangle \nonumber \\
&\times& \frac{1}{E-\mathcal{E}+i\eta} \langle\chi_{\mathcal{E}}^{(-)}(E)\mid 
\ . \label{eq.sperep}
\fineq

The hadron tensor can be reduced to a single-particle expression and can 
be written in an expanded form as 
\inieq
W^{\mu\nu}(\omega , q) &=& -\frac{1}{\pi} \sum_n  \textrm{Im} \bigg[
 \int_M^{\infty} \diff \mathcal{E} \frac{1}{E_{\textrm
{f}}-\varepsilon_n-\mathcal{E}+i\eta}  \nonumber \\
&\times&  T_n^{\mu\nu}(\mathcal{E} ,E_{\textrm{f}}-\varepsilon_n) \bigg]
\ , \label{eq.pracw}
\fineq
where $n$ denotes the eigenstate of the residual Hamiltonian of $A$ interacting
nucleons related to the discrete eigenvalue $\varepsilon_n$ and
\inieq
T_n^{\mu\nu}(\mathcal{E} ,E) &=& \lambda_n\langle \varphi_n
\mid j^{\nu\dagger}(\q) \sqrt{1-\mcv'(E)}
\mid\tilde{\chi}_{\mathcal{E}}^{(-)}(E)\rangle \nonumber \\
&\times& \!\! \langle\chi_{\mathcal{E}}^{(-)}(E)\mid  \sqrt{1-\mcv'(E)} j^{\mu}
(\q)\mid \varphi_n \rangle  \ . \label{eq.tprac}
\fineq
$\lambda_n$ is the spectral strength of the hole state 
 $\mid \varphi_n \rangle$, which is the normalized overlap integral between 
 $\mid\Psi_0\rangle$ and $\mid\ n \rangle$, while the factor 
 $\sqrt{1-\mcv'(E)}$ was shown in Ref.~\cite{ee} to account for
interference effects between different channels and to allow the replacement of
the mean field $\mcv$ by the phenomenological optical potential 
$\mcv_{\textrm L}$.   

After calculating the limit for $\eta \rightarrow +0$, 
Eq. (\ref{eq.pracw}) reads
\inieq
W^{\mu\nu}(\omega , q) = \sum_n \Bigg[ \textrm{Re} T_n^{\mu\nu}
(E_{\textrm{f}}-\varepsilon_n, E_{\textrm{f}}-\varepsilon_n)  \nonumber
\\
- \frac{1}{\pi} \mathcal{P}  \int_M^{\infty} \diff \mathcal{E} 
\frac{1}{E_{\textrm{f}}-\varepsilon_n-\mathcal{E}} 
\textrm{Im} T_n^{\mu\nu}
(\mathcal{E},E_{\textrm{f}}-\varepsilon_n) \Bigg] \ , \label{eq.finale}
\fineq
where $\mathcal{P}$ denotes the principal value of the integral. 
Eq. (\ref{eq.finale}) separately involves the real and imaginary parts of 
$T_n^{\mu\nu}$. 

The second matrix element in Eq. (\ref{eq.tprac}), with the inclusion of 
$\sqrt{\lambda_n}$ and disregarding the square root correction due to 
interference effects, is the transition amplitude for the single-nucleon 
knockout 
from a nucleus in the state $\mid \Psi_0\rangle$ leaving the residual nucleus 
in the state $\mid n \rangle$. The attenuation of its strength,
mathematically due to the imaginary part of $\mcv^{\dagger}$, is related to the
flux lost towards the channels different from $n$. In the inclusive response
this attenuation must be compensated by a corresponding gain, due to the flux
lost, towards the channel $n$, by the other final states asymptotically
originated by the channels different from $n$. In the description provided by
the spectral representation of Eq. (\ref{eq.sperep}), the compensation is
performed by the first matrix element in the right hand side of 
Eq. (\ref{eq.tprac}), where the imaginary part of $\mcv$ has the effect of 
increasing the strength. Similar considerations can be made, on the purely 
mathematical ground, for the integral of Eq. (\ref{eq.finale}), where the 
amplitudes involved in $T_n^{\mu\nu}$ have no evident physical meaning when 
${\mathcal{E}}\neq E_{\rm{f}}-\varepsilon_n$. 
We want to stress here that in the Green's function approach it is just the 
imaginary part of $\mcv$ which accounts for the redistribution of the strength 
among different channels.

The cross sections and the response functions of the inclusive QE
neutrino (antineutrino)-nucleus scattering are calculated from the 
single-particle expression of the hadron tensor in  Eq. (\ref{eq.finale}). 
After the replacement of the mean field $\mcv(E)$ by the empirical optical 
model potential $\mcv_{\textrm {L}}(E)$, the matrix elements of the nuclear 
current operator in Eq. (\ref{eq.tprac}), which represent the main ingredients 
of the calculation, are of the same kind as those giving the transition 
amplitudes  of the electron induced nucleon knockout reaction in the 
relativistic distorted wave impulse approximation (RDWIA)~\cite{meucci1}. 
Thus, the same treatment can be used which was successfully applied to 
describe exclusive $(e,e^{\prime}p)$ and $(\gamma,p)$ 
data~\cite{meucci1,meucci2}. Here, of course, the nuclear electromagnetic 
current must be replaced by the nuclear weak charged-current operator of 
Eq. (\ref{eq.cc}) .

The relativistic final wave function is written, as in Refs.~\cite{ee,meucci1},
in terms of its upper
component following the direct Pauli reduction scheme, i.e.,
\inieq
\!\!\!\! \chi_{\cal{E}}^{(-)}(E) \!\! = \!\! \left(\begin{array}{c} 
{\displaystyle \Psi_{\textrm {f}+}} \\ 
\frac{\displaystyle 1} {\displaystyle 
M+{\cal E}+S^{\dagger}(E)-V^{\dagger}(E)}
{\displaystyle \mbox{\boldmath $\sigma$}\cdot\p
        \Psi_{\textrm {f}+}} \end{array}\right) \ ,
\fineq
where $S(E)$ and $V(E)$ are the scalar and vector 
energy-dependent
components of the relativistic optical potential for a nucleon
with energy $E$ \cite{chc}. 
The upper component, $\Psi_{\textrm {f}+}$, is related to a two-component
spinor, $\Phi_{\textrm{f}}$, which solves a
Schr\"odinger-like equation containing equivalent central and 
spin-orbit potentials, obtained from the relativistic scalar and vector 
potentials \cite{clark,HPa}, i.e.,
\inieq
\Psi_{\textrm {f}+} &=& \sqrt{D_{{\cal E}}^{\dagger}(E)}\Phi_{\textrm{f}} \ , \\
D_{{\cal E}}(E) &=& 1 + \frac{S(E)-V(E)}{M+{\cal E}} \ , \label{eq.darw}
\fineq
where $D_{{\cal E}}(E)$ is the Darwin factor. 

As no relativistic optical potentials are available for 
the bound states, the wave functions $\varphi_n$ are taken as 
the Dirac-Hartree solutions of a relativistic Lagrangian
containing scalar and vector potentials \cite{adfx,lala}.

\section{Results}
\label{result}

The calculations have been performed with the same
bound state wave functions and optical potentials as in
Refs.~\cite{meucci1,meucci2,ee}, where the RDWIA  was able to 
reproduce $\left(e,e^{\prime}p\right)$, $\left(\gamma,p\right)$, and 
$\left(e,e^{\prime}\right)$ data. 

The relativistic bound state wave functions have been obtained from 
Ref.~\cite{adfx}, where relativistic Hartree-Bogoliubov equations are solved in
the context of a relativistic mean field theory and reproduce
single-particle properties of several spherical and deformed nuclei~\cite{lala}. 
The scattering state is calculated by means of
the energy-dependent and A-dependent EDAD1 complex phenomenological optical 
potential of Ref.~\cite{chc}, which is fitted to proton
elastic scattering data on several nuclei in an energy range up to 1040 MeV.

The initial states $\mid \varphi_n \rangle$ are taken as
single-particle one-hole states in the target. A pure shell model is 
assumed for the nuclear structure, i.e., we take a unitary spectral strength 
for each single-particle state and the sum runs over all the occupied states.

The results presented in the following contain the contribution of only
the first term in Eq. (\ref{eq.finale}). The calculation of the second 
term, which requires integration over all the eigenfunctions of the
continuum spectrum of the optical potential, is a complicate 
task. Its contribution has been estimated to be small in the kinematics 
explored; hence, it is neglected in the present calculations.

In order to show up the effect of the optical potential on the inclusive 
reaction, the results obtained in the present approach are compared with those 
given by different approximations. 

In the simplest approach the optical potential is neglected, i.e., 
$\mathcal{V} = \mathcal{V}^{\dagger} =0$ in Eq. (\ref{eq.op}), and the plane 
wave approximation is assumed for the final state wave functions ${\chi}^{(-)}$
and  $\tilde{\chi}^{(-)}$. In this plane wave impulse approximation (PWIA) FSI
between the outgoing nucleon and the residual nucleus are completely neglected. 

In another approach the imaginary part of the optical potential is neglected 
and only the real part is included. This approximation, that was sometimes used 
in the past, conserves the flux, but it is inconsistent with the exclusive 
process, where a complex optical potential must be used. Moreover, the use of 
a real optical potential is unsatisfactory from a theoretical point of view, 
since the optical potential has to be complex owing to the presence of open 
channels. 

\begin{figure}[h]
\includegraphics[height=12cm, width=9cm]{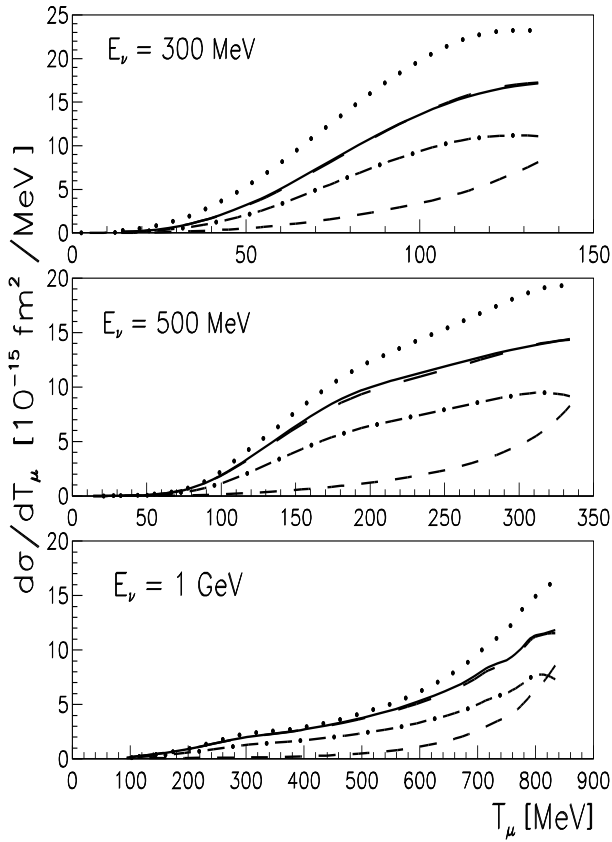} 
\vskip -0.3cm
\caption {The cross sections of the $^{16}$O$(\nu_{\mu},\mu^-)$ reaction, 
integrated over the muon angle, for $E_\nu$ = 300, 500, and 1000 MeV.  
Solid lines represent the result of the Green's function approach,
dotted lines give PWIA, long-dashed lines show the result with
a real optical potential, and dot-dashed lines the contribution of 
the integrated exclusive reactions with one-nucleon emission.
Short dashed lines give the cross sections of the 
$^{16}$O$(\bar\nu_{\mu},\mu^+)$ reaction calculated with the Green's 
function approach. }
\label{fig1}
\end{figure}
\begin{figure}[h]
\includegraphics[height=12cm, width=9cm]{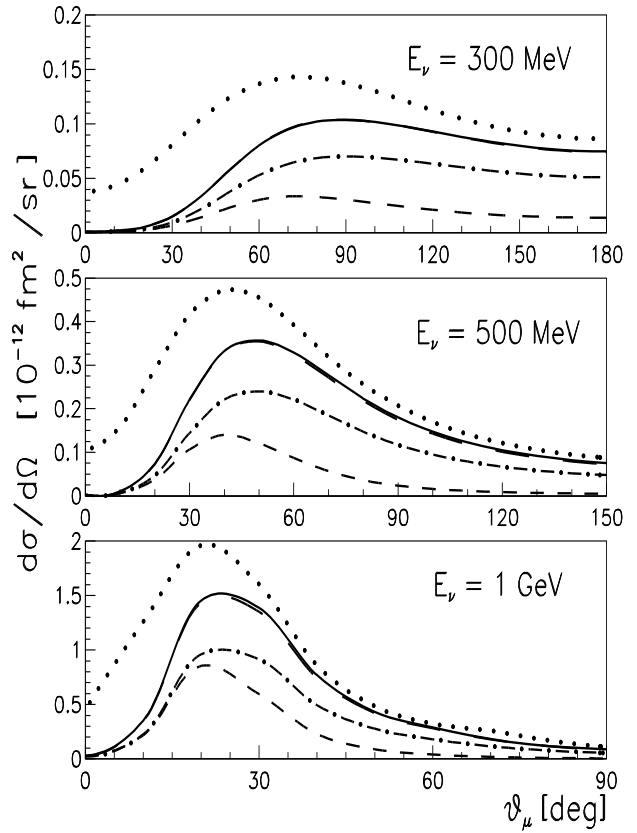} 
\vskip -0.3cm
\caption {The cross sections of the $^{16}$O$(\nu_{\mu},\mu^-)$ reaction, 
integrated over the muon energy, for $E_\nu$ = 300, 500, and 1000 MeV as a 
function of the scattering angle of the outgoing muon $\theta_\mu$.  
Line convention as in Fig. \ref{fig1}. }
\label{fig2}
\end{figure}
\begin{figure}[h]
\includegraphics[height=11cm, width=9cm]{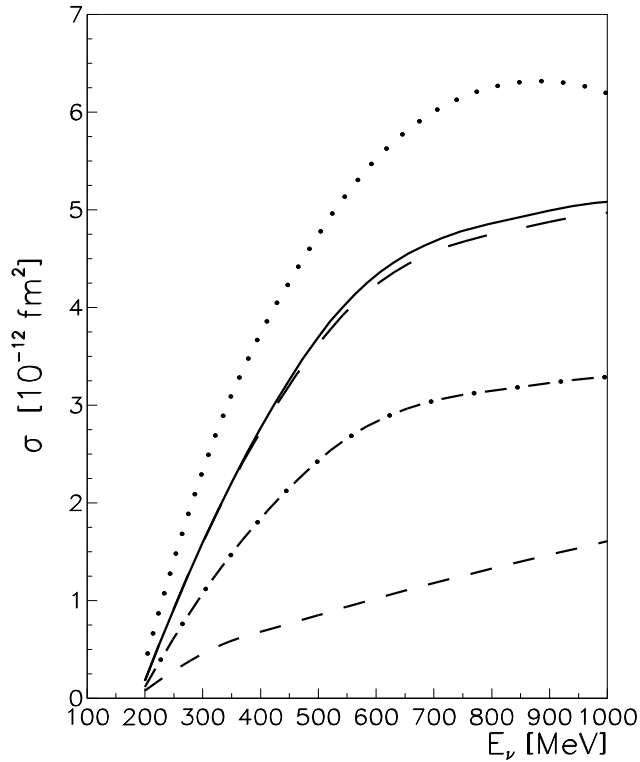} 
\vskip -0.3cm
\caption {The total cross section of the $^{16}$O$(\nu_{\mu},\mu^-)$ reaction, 
integrated over the muon energy and angle, in terms of the neutrino energy 
$E_\nu$.  
Line convention as in Fig. \ref{fig1}. }
\label{fig3}
\end{figure}

The partial contribution given to the inclusive process  
by the sum of all the integrated exclusive reactions with one-nucleon emission 
is also shown in the following for a comparison. 
In this case only the eigenfunctions ${\chi}^{(-)}$ 
of $\mathcal{V}^{\dagger}$ are included and the imaginary part of the potential 
produces an absorption which does not conserve the total flux. We note that in 
the Green's function approach of Eqs.~(\ref{eq.tprac}) and (\ref{eq.finale}) 
FSI are treated, consistently with the exclusive process, by means of a 
complex optical potential. The imaginary part, however, does not produce a 
global loss of flux and it is responsible for the redistribution of the 
strength among different channels. 

As a first study case we have considered the $^{16}$O target nucleus, for which
the adopted single-particle wave functions and optical potentials have
given a good agreement between RDWIA calculations and ($e,e'p$), ($\gamma,p$), 
and ($e,e'$) data.

In Fig. \ref{fig1} the cross sections of the $^{16}$O$(\nu_{\mu},\mu^-)$ 
reaction, integrated over the muon scattering angle, are displayed as a 
function of the muon kinetic energy $T_\mu$ for three different values of the 
incident neutrino energy $E_\nu$ = 300, 500 and 1000 MeV. 
The behaviour of the calculated cross sections is similar for the different 
energies. The effect of the optical potential increases with $T_\mu$ and 
decreases increasing $E_\nu$. At 300 MeV, the result of the PWIA is much higher 
than the one of the Green's function approach, while at 1 GeV the two results
are almost the same but at the highest values of $T_\mu$. 
The sum of the exclusive one-nucleon emission cross sections is always much 
smaller than the complete result. The difference indicates the relevance of
inelastic channels and is due to the loss of flux produced by the absorptive 
imaginary part of the optical potential. In contrast, the cross sections 
calculated  with only the real part of the optical potential are practically 
the same as the ones obtained with the Green's function approach. Although the 
use of a complex optical potential is conceptually important from a 
theoretical point of view, the negligible differences given by the two results 
mean that the conservation of flux, that is fulfilled in both calculations, is 
the most important condition in the present situation. In contrast, significant 
differences are obtained with a real optical potential in the inclusive 
electron scattering~\cite{capuzzi,ee}.

Qualitatively similar results are obtained in Fig. \ref{fig2},
where the cross sections integrated over the muon energy are displayed
for $E_\nu$ = 300, 500, and 1000 MeV as a function of the scattering angle 
of the outgoing muon.

The global effect of FSI is clearly shown in Fig. \ref{fig3}, where the cross 
sections are integrated over the energy and the angle of the outgoing muon. 
The PWIA result is always much larger, while the loss of flux produced by 
the absorptive optical potential in the exclusive processes produces a too 
small cross section. In contrast, an optical potential with only a real 
component seems able to give a result comparable with the one of the Green's 
function approach. Small differences are found only at higher neutrino energies.

In Figs. \ref{fig1}, \ref{fig2}, and \ref{fig3} also the cross sections for 
the $^{16}$O$(\bar\nu_{\mu},\mu^+)$ reaction are shown for a comparison. 
They are always much smaller than the corresponding cross sections with an
incident neutrino.

\begin{figure}[h]
\includegraphics[height=11cm, width=9cm]{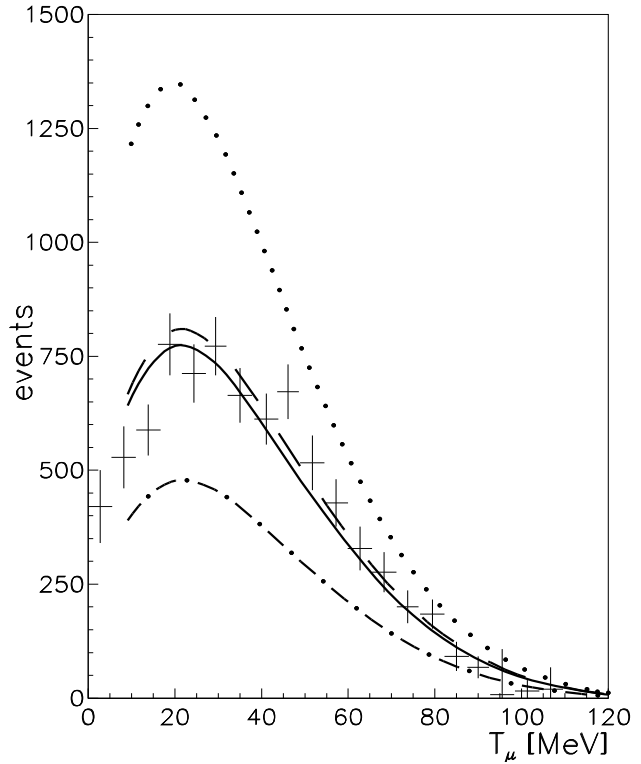} 
\vskip -0.3cm
\caption {The distribution of the muon kinetic energy for the inclusive 
$^{12}$C$(\nu_{\mu},\mu^-)$ reaction.
Experimental data from Ref.~\cite{athan}. The result of the Green's 
function approach is normalized to the experimental data. The other 
curves are scaled, accordingly.
Line convention as in Fig. \ref{fig1}. }
\label{fig4}
\end{figure}

The different approaches have been compared with the experimental results 
of the LSND collaboration at Los Alamos for the $^{12}$C$(\nu_{\mu},\mu^-)$ 
reaction~\cite{albert,athan,auer}. The calculations have been flux-averaged 
over the Los Alamos neutrino spectrum. In Fig. \ref{fig4} the Green's 
function approach is normalized to the experimental data, while the other
results are accordingly scaled. The shape of experimental data is
reasonably reproduced. The flux-averaged cross section integrated over the 
muon energy gives 11.15 $\times$ 10$^{-40}$ cm$^2$. At the low energy of the 
experiment, however, other processes, which are here not included, besides the 
quasielastic scattering can affect the inclusive reaction, in particular the 
excitation of the discrete states of $^{12}$C. The experimental 
value (10.6 $\pm$ 0.3 $\pm$ 1.8) $\times$10$^{-40}$ cm$^2$~\cite{auer}
is slightly overestimated. 
The results obtained by other 
calculations~\cite{min,singh,hayes,volpe,kolbe,jacho,maieron} give larger 
values.

\section{Summary and conclusions}
\label{conc}

We have applied to $(\nu_{\mu},\mu^-)$ and $(\bar\nu_{\mu},\mu^+)$ reactions 
an  approach based on the spectralization of the single-particle 
Green's function in terms of the eigenfunctions of the complex optical 
potential and of its Hermitian conjugate. This approach has proved to be rather 
successful in describing inclusive electron scattering. Its advantage stands 
in the fact that it is able to include in a simple way and keeping flux 
conservation the final state interactions by using an optical potential 
which is essential to reproduce exclusive electron knockout reactions.

The method is applied within a relativistic framework to weak charged-current 
reactions for an energy up to 1 GeV, where nuclear structure effects are 
important and dominant with respect to nucleon-resonance excitations.
The reaction mechanism is assumed to be a direct one, where the
incident neutrino interacts with only one neutron and a proton is emitted
together with a negative-charged muon. A single-particle model is used to
describe the structure of the nucleus and a sum over all single-particle 
occupied states is performed. 

Calculations for the $^{16}$O target nucleus have been presented at neutrino 
energies up to 1 GeV. The optical potential and flux conservation have a large 
effect on the cross sections. 
For $^{12}$C, the results averaged over the experimental flux of neutrinos 
are compared with the available data. 
A fair agreement is obtained in arbitrary units. The integrated cross section 
results somewhat larger than the experimental value.

\begin{acknowledgments}

We would like to thank W.M. Alberico for useful discussions.

\end{acknowledgments}



\end{document}